\documentstyle{mn}
\input{epsf}

\title{The inverse problem for pulsating neutron stars: A 
``fingerprint analysis'' for the supranuclear equation of state}

\author[K. D. Kokkotas, T. A. Apostolatos and  N. Andersson]
{K. D. Kokkotas$^{1,2}$, T. A. Apostolatos$^{3}$ and
N. Andersson$^{4}$\\ 
$^{1}$ Department of Physics, Aristotle University of Thessaloniki,
       Thessaloniki 54006, Greece.\\
$^{2}$ Max Planck Institute for Gravitational Physics, 
       The Albert Eistein Institute,
       D-14473 Potsdam, Germany. \\
$^{3}$ Department of Physics, University of Athens, Greece. \\
$^{4}$ Department of Mathematics, University of Southampton,
Southampton, UK.\\
}

\date{Accepted 1999  (?).
      Received 1999  (?);
      in original form  1999}
\begin{document}
\maketitle

\begin{abstract}
We study the problem of detecting, and infering astrophysical
information from, gravitational waves from a pulsating neutron star. We show
that the fluid $f$ and $p$-modes, as well as the gravitational-wave $w$-modes
may be detectable from sources in our own galaxy, and investigate how
accurately the  frequencies and damping rates  of these modes can be infered
from a noisy gravitational-wave data stream. Based on the conclusions of this
discussion we propose a strategy for  revealing the supranuclear equation of
state using the neutron star fingerprints: the observed frequencies of an $f$
and a $p$-mode. We also discuss how well the source can be located in the sky
using observations with several detectors. 

\end{abstract}


\section {Introduction}

As we approach the next millennium there is a focussed worldwide
effort to construct devices that will enable the first undisputed
detection of gravitational waves. A network of large-scale
ground-based laser-interferometer detectors (LIGO, VIRGO, GEO600,
TAMA300) is under construction, while the sensitivity of the
several resonant mass detectors that are already in operation continues to
be improved. At the present time it seems likely that the
gravitational-wave window to the Universe will be opened within
the next five to ten years, and that gravitational-wave astronomy
will finally become a reality.

An integral part in this effort
is played by theoretical modelling of the expected sources.
Theorists are presently racking their brains to think of various
sources of gravitational waves that may be observable once the new
ultrasensitive detectors operate at their optimum level, and of
any piece of information one may be able to extract from such
observations.
One of the most challenging goals that can (at least, in principle)
be achieved via gravitational-wave detection is the determination of
the equation of state of matter at supranuclear densities.

We have recently argued that observed gravitational waves from
the various nonradial pulsation modes of a neutron star can be
used to infer both the mass and the radius of the star with
surprisingly good accuracy, and thus put useful constraints on the
equation of state \cite{ak96,ak98}. This conclusion was, however, based
on an ``ideal'' detection situation that ignored the various
uncertainties associated with the analysis of a noisy data-stream.
In the present paper we analyze thoroughly this idea by
incorporating all possible statistical errors that might arise in
estimating the parameters associated with the neutron star equation of
state.

The spectrum of a pulsating relativistic star is known to be
tremendously rich, but most of the associated pulsation modes
are of little relevance for gravitational-wave detection.
From the gravitational-wave point of view one would expect
the most important modes
to be the fundamental ($f$) mode of fluid oscillation,
the first few pressure ($p$) modes
and the first gravitational-wave ($w$) modes \cite{ks92}.
For details on
the theory of relativistic
stellar pulsation we refer the reader to a recent review article by
one of us \cite{kokkotas97}. That the bulk of the energy from an
oscillating neutron star is, indeed,
radiated through these modes has been demonstrated by
numerical experiments \cite{aaks,2ns}.

Recently, two of us \cite{ak98} provided data for the relevant
pulsation modes of several stellar models for each of twelve
proposed realistic equations of state, thus extending earlier results
of Lindblom and Detweiler \shortcite{ld83}. This
data was then used to obtain empirical relations between the observables
(frequency and damping time) of the $f$- and the $w$-modes and
the stellar parameters (mass and radius). It was shown
that these relations could be used to infer both the radius and
the mass of the star (typically with an error smaller than 10\%),
i.e., to take the fingerprints of the star. It was also pointed out
that, since no general empirical relations could be inferred for
the $p$-modes, they could prove important for deducing the actual
equation of state once the radius and the mass of the star is
known. 
The empirical relations, including the relevant statistical errors,
are listed in Appendix~1.

The proposed strategy can potentially be of great importance to
gravitational-wave astronomy, since most stars are expected to
oscillate nonradially. The evidence for this is compelling:  Many $p$-modes
(as well as possible $g$-modes and $r$-modes) have been observed
in the sun and there are strong indications that similar modes are
excited also in more distant stars. In principle, one would expect
the modes of a star to be excited in any dynamical scenario that
leads to significant asymmetries. Still, one can only hope to
observe gravitational waves from the most compact stars. Hence,
our attention is restricted to neutron stars (or possible strange
stars \cite{alcock}, if they exist). Furthermore, to lead to
detectable gravitational waves the modes must be excited to quite
large amplitudes, which means that only the most violent processes are of interest.

There are several scenarios in which the various pulsation modes
may be excited to an interesting level: 
(1) A supernova explosion
is expected to form a wildly pulsating neutron star that emits
gravitational waves. The current estimates for the energy radiated
as gravitational waves from supernovae is rather pessimistic,
suggesting a total release of the equivalent to $10^{-6} M_\odot
c^2$, or so. However, this may be a serious underestimate if the
gravitational collapse in which the neutron star is formed is
strongly non-spherical. Optimistic estimates suggest that as much as
$10^{-2}M_\odot c^2$ may be released in extreme events.
(2) Another potential excitation mechanism
for stellar pulsation is a starquake, e.g., associated with a pulsar
glitch. The typical energy released in this process may be of the
order of the maximum mechanical energy that can be
stored in the crust, estimated to be at the level of $10^{-9}-10^{-7}M_\odot
c^2$ \cite{blaes,mock}. This is also an interesting possibility  considering 
the recent conclusion that the soft-gamma repeaters are likely to be so-called
magnetars,  neutron stars with extreme magnetic fields \cite{dt92}, that
undergo frequent starquakes. It seems very likely that some pulsation modes are
excited by the rather dramatic events that lead to the most energetic bursts 
seen from these systems. Indeed, Duncan \shortcite{duncan} has recently
argued that toriodal modes in the crust should be excited. If modes are excited
in these systems, an indication of the energy released in the most powerful
bursts is the  $10^{-9}M_\odot
c^2$  estimated for the March 5 1979 burst in SGR~0526-66. The maximum energy
should certainly not exceed the total supply in the magnetic field $\sim
10^{-6} (B/10^{15} G)^2 M_\odot c^2$ \cite{dt92}. The possibility
that a burst from a soft gamma-ray repeater may have a gravitational-wave
analogue is very exciting. (3) The coalescence of two neutron stars at the end
of binary inspiral may form a pulsating remnant. It is, of course, most likely,
that a black hole is formed when two neutron stars coalesce, but even in that
case the eventual collapse may be halted long enough (many dynamical
timescales) that several oscillation modes could potentially be identified
\cite{baum}.  Also, stellar oscillations can
be excited by the tidal fields of the two stars during the
inspiral phase that preceeds the merger \cite{KS95}.   (4) The star may
undergo a dramatic phase-transition that leads to a mini-collapse. This would
be the result of a sudden softening of the equation of state (for example,
associated with the formation of a  condensate consisting of pions or kaons). A
phase-transition could lead to a sudden contraction during which a
considerable part of the stars gravitational binding energy would be released,
and it seems inevitable that part of this energy would be channeled into
pulsations of the remnant. Large amounts of energy that could be released in
the most extreme of these scenarios: A contraction of (say) 10\% can easily
lead to the release of $10^{-2}M_\odot c^2$. Transformation of a neutron star
into a strange star is likely to induce pulsations in a similar fashion.

It is reasonable to assume that the bulk of the total energy of the
oscillation is released through a few of the stars quadrupole pulsation
modes in all these scenarios. We will assume that this is the case and
assess the likelihood that the associated gravitational waves will
be detected. Having done this we discuss the inverse problem, and
investigate how accurately the neutron star parameters can be
inferred from the gravitational wave data.

Before we proceed with our main analysis it is worthwhile making
one further comment. In this study we neglect the effects of
rotation on the pulsation modes. This is not because rotation
plays an insignificant role. On the contrary, we expect rotation
to be highly relevant in many cases. The recent discovery that the
so-called $r$-modes may be unstable \cite{and,fm} and lead to rapid spin
down of a young neutron star that is born rapidly rotating \cite{aks},
highlights the fact that it is not sufficient to consider only
non-rotating stars. Still, as far as most pulsation modes are
concerned, one would expect rotation to have a significant effect
only for neutron stars with very short period, and the present study
may well be reasonable for stars with periods longer than, say, 20
ms. Furthermore, it has been argued that neutron stars are
typically born slowly rotating \cite{spruitt}. If that
is the typical case, then our results could be relevant for most
newly born neutron stars. A more pragmatic reason for not
including rotation in the present study is that detailed
data for modes of rotating neutron stars is not yet available.
Once such data has been computed the present study should be
extended to incorporate rotational effects.

Having pointed out this caveat, we are prepared to proceed with
the discussion of the present results. The rest of the paper is
organized as follows. In Section 2 we basically repeat the
analysis of Finn \shortcite{finn92}, in a slightly different form
in order to compute the measurement errors of frequency and
damping time of a gravitational wave that is emitted from a
pulsating neutron star, and extend this analysis, so as to compute
the accuracy by which one can estimate various parameters of the
star. In Section 3 we transform our results into a form that shows
clearly how plausible it is to determine the equation of state of
a neutron star by analyzing the gravitational wave data. In Section
4 we discuss how detection of mode signals can be used to reveal
the nuclear equation of state. Section 5 is a brief discussion of
how accurately one can expect to be able to locate the source in
the sky. The final section presents our conclusions. Appendix~1
contains empirical relations for oscillation frequency and damping rate 
of the $f$ and $w$-modes in terms of the stellar parameters (mass and radius), 
deduced from twelve realistic equations of state \cite{ak98}. Appendix~2 lists
the elements of the Fisher and covariance matrices used in the statistical
analysis of the mode-signals.

\section {Statistical analysis of observed mode-signals}

Suppose that one tries to detect the gravitational waves
associated with the stellar pulsation modes that are excited when
(say) a neutron star forms after a supernova explosion. Since all 
modes are relatively short lived, the detection situation is
similar to that for a perturbed rotating black hole
\cite{Ech,finn92}. For each individual mode the signal is
expected to have the following form:
\begin{equation}
h(t)=
\cases{0                                  & for $t<T$, \cr
       {\cal A} e^{- (t-T)/ \tau} \sin[2 \pi f (t-T)] & for $t>T$. \cr}
\label{template}
\end{equation}
Here, $\cal A$ is the initial amplitude of the signal, $T$ is 
its arrival time, and $f$
and $\tau$ are the frequency and damping time of the oscillation, respectively.
Since the violent formation of a neutron star is a very complicated
event, the above form of the waves becomes realistic only at the
late stages when the remnant is settling down and its pulsations
can be accurately described as a superposition of the various
modes, either fluid or spacetime ones, that have been excited. At
earlier times ($t<T$) the waves are expected to have a random
character that is completely uncorrelated with the intrinsic noise
of an earth-bound detector. This partly justifies our
simplification of setting the waveform equal to zero for $t<T$.

The energy flux $F$ carried by any weak gravitational wave $h$ is
given by
\begin{equation}
F= {c^3 \over 16 \pi G} | {\dot h} |^2 \;, \label{flux}
\end{equation}
where $c$ is the speed of light and $G$ is Newton's gravitational
constant. Thus, when gravitational waves emitted from a pulsating
neutron star hit such a detector on Earth, their initial amplitude
will be \cite{Tho,Schutz}
\begin{eqnarray}
{\cal A} \sim 2.4 &\times& 10^{-20} \left( {E_{\rm gw}  \over
10^{-6} M_{\odot} c^2}  \right)^{1/2} \nonumber \\ &\times& \left(
{10 {\rm kpc}  \over r                       }  \right) \left( {1
{\rm kHz}  \over f                       }  \right) \left( {1 {\rm
ms}   \over \tau                    } \right)^{1/2}
\label{strength}
\end{eqnarray}
where $E_{\rm gw}$ is the energy released through the mode and $r$ is
the distance between detector and source. Again, $f$ is the frequency of
gravitational waves and $\tau$ is the damping time: the rate
at which the amplitude of the mode decays as energy is carried
away from the star.

In order to reveal this kind of  signal from the noisy output of a
detector one could use templates of the same form as the expected signal
(so called matched filtering).
Following the analysis of Echeverria \shortcite{Ech} the
signal-to-noise ratio is found to be
\begin{equation}
\left( {S \over N} \right)^2 = \rho^2 \equiv
2 \langle h \mid h \rangle =
{4 Q^2 \over 1+4 Q^2} \; {{\cal A}^2 \tau \over 2 S_n} \;,
\label{SNR}
\end{equation}
with
\begin{equation}
Q \equiv \pi f \tau \;,
\label{Qdef}
\end{equation}
being the quality factor of the oscillation, and
$S_n$ the spectral density of the detector (assumed to be constant
over the bandwidth of the signal).

In Eq.~(\ref{SNR}), we used the
following definition for the scalar product between two functions:
\begin{eqnarray}
\langle h_1 \mid h_2 \rangle
&\equiv&
\int_{-\infty}^{\infty} df
{\tilde h_{1}(f) \tilde h_{2}^{\ast}(f)
\over
S_{n}( \mid f \mid )}
=
{1 \over S_n}
\int_{-\infty}^{\infty} df
{\tilde h_{1}}(f) {\tilde h_{2}}^{\ast}(f) \nonumber \\
&=&
{1 \over S_n}
\int_{-\infty}^{\infty} dt
h_{1}(t) h_{2}(t) \;.
\label{product}
\end{eqnarray}
To compute the accuracy by which the parameters of the signal can be
determined, we first define the dimensionless parameters
{$\epsilon,\eta,\xi,\zeta$} as
\begin{eqnarray}
f_{\rm o}        \epsilon &\equiv& f        - f_{\rm o}        \cr
\tau_{\rm o}     \eta     &\equiv& \tau     - \tau_{\rm o}     \cr
\tau_{\rm o}     \zeta    &\equiv& T        - T_{\rm o}        \cr
{\cal A}_{\rm o} \xi      &\equiv& {\cal A} - {\cal A}_{\rm o} \;,
\end{eqnarray}
where $f_{\rm o},\tau_{\rm o},T_{\rm o},{\cal A}_{\rm o}$ are the
true values of the four quantities $f,\tau,T,{\cal A}$. These new
parameters are simply relative deviations of the measured
quantities from their true values. One can then construct the
Fisher information matrix $\Gamma_{i j}$ and by inverting it,
obtain all possible information about the measurement accuracy of
each parameter of the signal, and the correlations between the
errors of the parameters. The components of the symmetric Fisher
matrix, which is defined by
\begin{equation}
\Gamma_{i j} \equiv
2 \Big\langle
{\partial h \over \partial \theta_i} \mid
{\partial h \over \partial \theta_j}
\Big\rangle \;,
\label{Fisher1}
\end{equation}
where $\theta_i=(\epsilon, \eta,\zeta,\xi)$ are the parameters of
the signal, are listed in Appendix~2.

The inverse of the Fisher matrix, $\Sigma_{i j} \equiv \Gamma_{i
j}^{-1}$, the so-called covariance matrix, is the most important
quantity from the experimental point of view. Its components, that
are directly related with the measurement errors of the
parameters, are also listed in Appendix~2. The purpose of the
present analysis is to try to identify the nuclear equation of
state from supposedly detected mode data. The  relevant parameters
for this analysis are the frequency $f$ and damping time $\tau$.
Hence, we only need $\Sigma_{\epsilon \epsilon}$ and $\Sigma_{\eta
\eta}$  ---the squares of the relative errors of $f$ and $\tau$
respectively--- from Appendix~2. We will also discuss the
possibility that the signal will be seen by several detectors. If
that is the case, we can use the time of arrival $T$ to locate the
position of the source in the sky. For this discussion we need
$\Sigma_{\zeta \zeta}$, see Section~5.

The estimated errors of the actual measurements should be taken
into account along with the statistical errors in our empirical
relations (cf. Appendix~A) in order to compute the parameters of
the pulsating star. The law of error propagation will be employed
to estimate the errors of the parameters of the star: When one
tries to estimate the value of a quantity $z$ which is given as a
function of other quantities $z=f(x_1,x_2,\cdots)$ that can be
measured directly, the error that should be attributed to the
former one is given by
\begin{equation}
\sigma_{z}=
\left[
\sum_{i}   \left( {\partial f \over \partial x_i} \sigma_{x_i} \right)^2 +
\sum_{i \neq j} \left( {\partial f \over \partial x_i}
                  {\partial f \over \partial x_j}
                   \sigma_{x_i} \sigma_{x_j} r_{x_i x_j} \right)
\right] ^{1/2} \;,
\label{errorpropagation}
\end{equation}
where $\sigma_{x_i}$ is the measurement error of the quantity
$x_i$ and $r_{x_i x_j}$ is the correlation between the errors of
$x_i$ and $x_j$. In Section~4 we will use the measured values of
the frequencies of the $f$ and the first $p$-mode 
($f_f$ and $f_p$) to obtain information about the parameters of the
star. Since these are independent quantities (they are measured
separately by different templates), the second sum of the above
law will not be present in our calculations. The magnitude of the
estimated errors of neutron star parameters will determine the
efficiency of the method and our chance to achieve our final goal,
the determination of the equation of state of a neutron star.

\section {Detecting pulsation modes}

\subsection{Are the modes detectable?}

Two separate questions must be addressed in any discussion of
gravitational-wave detection. The first one concerns identifying a
weak signal in a noisy detector, thus establishing the presence of
a gravitational wave in the data. The second question regards
extracting the detailed parameters of the signal, e.g., the
frequency and e-folding time of a pulsation mode. To address
either of these issues we need an estimate of the spectral noise
density $S_n$ of the detector.

\begin{figure}
\centerline{\epsfxsize=9cm \epsfysize=9cm \epsfbox{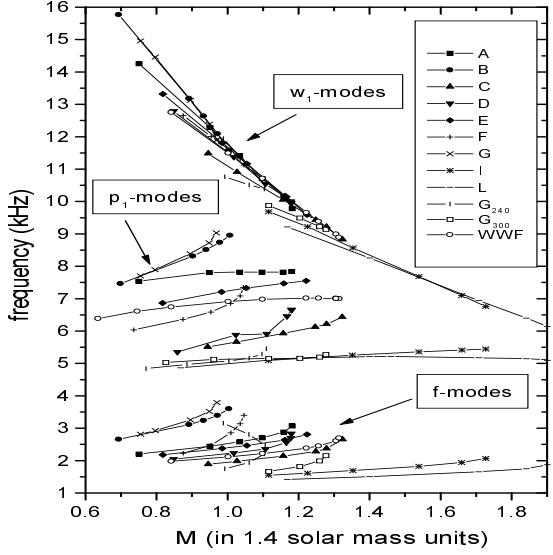}}
\caption{ This diagram shows the range of mode-frequencies for
twelve different equations of state, as functions of the normalized
mass (${\bar M}$ of the star. } \label{fig1}
\end{figure}

The few pulsation modes of a neutron star that may be detectable
through the associated gravitational waves all have rather high
frequencies; typically of the order of several kHz. To illustrate
this we show the mode-frequencies for all models considered by
Andersson and Kokkotas \shortcite{ak98} as a function of the
stellar mass in Figure~\ref{fig1}. From
this figure we immediately see that a detector must be sensitive
to frequencies of the order of 8-12 kHz and above to observe most $w$-modes.
Of course, it is also clear that some equations of state yield $w$-modes
at somewhat lower frequencies. 
For example, for massive neutron stars with
$M\approx 1.8-2.3 M_\odot$ (stiff EOS), as have been suggested 
for low-mass X-ray binaries, the $w$-mode frequency could be as 
low as 6~kHz (see also recent results for the axial $w$-modes by 
Benhar et al \shortcite{BBF99}).
The $p$-modes lie mainly in the range 4-8 kHz, while all $f$-modes
have frequencies lower than 4 kHz. This means that the
mode-signals we consider lie in the regime where an
interferometric detector is severely limited by the photon shot
noise. For this reason a detection strategy based on resonant
detectors (bars, spheres or even networks of small resonant
detectors \cite{frasca}) or laser interferometers operating in
dual recycling mode seems the most promising. In fact, the range of
mode-frequencies in Fig.~\ref{fig1} should motivate detailed studies
of the prospects for construction dedicated ultrahigh frequency
detectors. In the
following we will compare three different detectors:  The initial and advanced
LIGO interferometers, for which  \begin{equation}
S_n^{1/2} \approx h_m \left( { f \over \alpha f_m} \right)^{3/2}
{1\over \sqrt{f}}\ \mbox{Hz}^{-1/2} \;,
\end{equation}
with $h_m=3.1\times10^{-22}$, $\alpha=1.4$ and $f_m=160$~Hz for the
initial configuration, and
$h_m=1.4\times10^{-23}$, $\alpha=1.6$ and $f_m=68$~Hz respectively, 
for the advanced configuration   \cite{flanagan}. 
We also consider an ``ideal'' detector that is tuned to the frequency of the
mode and has sensitivity of the order of $S_n^{1/2} \approx 10^{-23}$
Hz$^{-1/2}$ 
(this is the sensitivity goal of detectors under
construction ). 
It should be noted that the Advanced LIGO estimates are roughly
valid also for spherical detectors such as TIGA, cf. Harry, Stevenson and
Paik \shortcite{harry}.

The detectability of the $f$, $p$ and $w$-modes for different
detectors can be assessed from (\ref{SNR}). The main problem in
doing this is the lack of realistic simulations providing
information about the level of excitation of various modes in
an astrophysical situation. Still, given the frequency and damping
rate of a specific mode we can ask what amount of energy must be
channeled through the mode in order to be detectable by a
given detector. We immediately find that 
detection of pulsating neutron stars from outside our own galaxy
is very unlikely. Let us consider a ``typical'' stellar model for
which the $f$-mode has parameters $f_f=2.2$~kHz and $\tau_f=0.15$~s.
This corresponds to a $1.4M_\odot$ neutron star according to the
Bethe-Johnson equation of state, cf. Table~3 of Andersson and Kokkotas
\shortcite{ak98}. For this example we find that the $f$-mode in the
Virgo cluster (at 15 Mpc) must carry an energy equivalent to more
than $0.3 M_\odot c^2$ to lead to a signal-to-noise ratio of 10 in
our ideal detector. Given that the total energy estimated to be
radiated as gravitational waves in a supernova is at the level of
$10^{-5}-10^{-6} M_\odot c^2$, we cannot realistically expect to
observe mode-signals from far beyond our own galaxy.

This means that the number of detectable events may be rather low.
Certainly, one would not expect to see a supernova in our galaxy
more often than once every thirty years, or so. Still, there are a
large number of neutron stars in our galaxy, all of which may be
be involved in dramatic events (see the introduction for some
possibilities) that lead to the excitation of pulsation modes.
The energies required to make each mode detectable (with a
signal-to-noise ratio of 10) from a source at the center of our galaxy (at 10
kpc) are listed in Table~\ref{tab1}. In the table we have used the
data for the ``typical'' stellar model, for which the characteristics
of the $f$-mode were given
above, $f_p=6$~kHz and $\tau_p=2$~s, and $f_w=11$~kHz and
$\tau_w=0.02$~ms. This data indicates that, even though the event
that excites the modes must be violent, the energy required to
make each mode detectable is not at all unrealistic. In fact, the
energy levels required for both the $f$- and $p$-modes are such
that detection of violent events in the life of a neutron star
should be possible, given the Advanced LIGO detectors (or
alternatively spheres with the sensitivity proposed for TIGA). On
the other hand, detection of $w$-modes with the broad band
configuration of LIGO seems unlikely. 
Detection of these modes,
which would correspond to observing a uniquely relativistic
phenomenon, requires dedicated high frequency detectors operating
in the frequency range above 6 kHz. Still, we believe that the data in
Table~\ref{tab1} illustrates that neutron star pulsation modes may
well be detectable from within our galaxy, and that the first
detection may in fact come as soon as the first generation of LIGO
detectors come on line.

\begin{table}
  \caption{The estimated energy (in units of $M_\odot c^2$)
  required in each mode in order to lead to a
  detection with signal-to-noise ratio of 10 from a pulsating neutron star at
  the center of our galaxy (10 kpc).}
  \begin{tabular}{@{}llll}
 Detector  & $f$-mode & $p$-mode & $w$-mode \\
Initial LIGO  & $4.9\times10^{-5}$ & $4.0\times 10^{-3}$ &  $6.8\times
10^{-2}$ \\ Adv LIGO  & $8.7\times10^{-7}$ & $7.0\times 10^{-5}$ &
$1.2\times 10^{-3}$ \\ Ideal & $1.4\times10^{-7}$ & $1.3\times
10^{-6}$ &  $6.4\times 10^{-6}$ \\
\end{tabular}
\label{tab1}\end{table}

\subsection{How well can we determine the mode parameters?}

Let us now discuss the precision with which we can hope to infer
the details of each pulsation mode. After inserting
Eq.~(\ref{strength}) into formulae (\ref{sigma11}) and
(\ref{sigma22}) we can compute the relative measurement error in
the frequency and the damping time of the waves by some
appropriately designed detector. After introducing a convenient
parameter ${\cal P}$ according to
\begin{equation}
{\cal P}^{-1} = \left( {S_{n}^{1/2}   \over 10^{-23}  {\rm
Hz^{-1/2}}} \right) \left( {r             \over 10  {\rm kpc} }
\right) \left( {E_{\rm gw}    \over 10^{-6}   M_{\odot} c^2}
\right)^{-1/2} \;,
\end{equation}
we find that the error estimates take the following form
\begin{equation}
{\sigma_f \over f} \simeq 0.0042 \  {\cal P}^{-1} \ {\sqrt
{1-2Q^2+8Q^4 \over 4Q^4}} \ \left( {\tau           \over 1  {\rm
ms}           }   \right)^{-1}   , \label{relerrorf}
\end{equation}
and
\begin{equation}
{\sigma_\tau \over \tau} \simeq 0.013 \ {\cal P}^{-1} \ {\sqrt
{10+8Q^2  \over Q^2}}    \ \left(  {f            \over 1  {\rm
kHz}           }   \right)        . \label{relerrortau}
\end{equation}
Also, for the time of arrival of the gravitational wave signal we
get from (\ref{sigma33})
\begin{equation}
\sigma_T \simeq 0.0042 \ {\cal P}^{-1} {\rm ms}         \ .
\label{relerrortime}
\end{equation}

To illustrate these results we show the relative errors associated
with the parameter extraction for the ``typical'' $1.4M_\odot$
stellar model we used in the previous section; see Table~\ref{tab2}. 
We assume that each
mode carries the energy required for it to be observed with
signal-to-noise ratio of 10, cf.  Table~\ref{tab1}. (This is a
convenient measure since it is independent of the particulars of
the detector.)

\begin{table}
  \caption{The relative errors in the extraction of the mode
  parameters assuming a signal-to-noise ratio of 10.}
  \begin{tabular}{@{}llll}
 & $\sigma_f/f$ & $\sigma_\tau/\tau$ & $\sigma_T$~($10^{-6}$~s) \\
$f$-mode & $8\times10^{-5}$ & $0.2$ & 11.3
\\ $p$-mode & $3\times10^{-6}$ & $0.2$ & 3.8
\\ $w$-mode & $0.1$ & $0.3$ & 1.7 \\
\end{tabular}
\label{tab2}\end{table}

From the sample data in Table~\ref{tab2} one sees clearly that,
while an extremely accurate determination of the frequencies of both the $f$-
and the $p$-mode is possible, it would be much harder to infer
their respective damping rates. It is also clear that an accurate
determination of both the $w$-mode frequency and damping will be
difficult. To illustrate this result in a different way, we can
ask how much energy must be channeled through each mode in order
to lead to a 1\% relative error in the frequency or the damping
rate, respectively. Let us call the corresponding energies $E_f$ and $E_\tau$.
This measure will then be detector dependent, so we list the
relevant estimates for the three detector configurations used
in Table~\ref{tab1}. When the data is viewed in this way, cf. Table~\ref{tab3},
we see that an accurate extraction of $w$-mode data will not be
possible unless a large amount of energy is released through these
modes. Furthermore, one would clearly need a detector that is
 sensitive at ultrahigh frequencies. In other words, it
seems unlikely that we will be able to use the $w$-modes to infer
the detailed neutron star parameters as we have previously
suggested \cite{ak96}.

\begin{table*}
 \centering
 \begin{minipage}{140mm}
 \caption{The estimated energy (in units of $M_\odot c^2$)
  required in each mode in order to lead to a relative error of
  1\% in the inferred mode-frequency ($E_f$) and damping rate
  ($E_\tau$). In cases where no entry is given, the required energy is
  unrealistically high (typically larger than $M_\odot c^2$).
  The distance to the source is assumed to be 10 kpc.}
  \begin{tabular}{@{}lllllll}
 Detector  & \multicolumn{2}{c}{$f$-mode} &  \multicolumn{2}{c}{$p$-mode} &
 \multicolumn{2}{c}{$w$-mode} \\
  & $E_f$ & $E_\tau$   & $E_f$ & $E_\tau$   & $E_f$ & $E_\tau$ \\
LIGO 1 & $3.1\times10^{-9}$ & $1.9\times 10^{-2}$ &  $2.8\times
10^{-10}$  & --- & --- & --- \\ Adv LIGO  & $5.5\times10^{-11}$ &
$3.5\times 10^{-4}$ & $4.9\times 10^{-12}$ & $2.8\times10^{-2}$ &
$0.17$ & ---
\\ Ideal & $8.8\times10^{-12}$ & $5.6\times 10^{-5}$ &
$8.8\times 10^{-14}$& $5\times10^{-4}$ & $9\times 10^{-4}$ &
$6.1\times 10^{-3}$  \\
\end{tabular}
\end{minipage}
\label{tab3}
\end{table*}

\section{Revealing the equation of state}

In the previous section we discussed issues regarding the
detectability of a mode-signal, and the accuracy with which the
parameters of the mode could be inferred from noisy gravitational
wave data. Let us now assume that we have detected the mode and
extracted the relevant parameters. We then naturally want to constrain
the supranuclear equation of state by deducing the
mass and the radius of the star (or combinations of them). In
principle, the mass and the radius can be deduced from any two
observables, cf., Table~2 of Andersson and Kokkotas
\shortcite{ak98}. In the absence of detector noise, several
combinations look promising, but in reality only few
combinations are likely to be useful.

Consider the following example: We could in principle deduce the
mass and the radius from a detected $f$-mode (assuming that we have
extracted both its frequency and damping rate). However, as can be seen
from  Table~\ref{tab2},  the  estimated relative error in frequency
is about three orders of magnitude smaller than the relative error
in damping time. Hence, if these two measurements are to be
used to determine the mass and the radius of the pulsating star
one must keep in mind that the measurement of frequency is far more
accurate than the measurement of damping time. And it is clear
that by combining $f_f$ with $\tau_f$ we will only get accurate
estimates for $M$ and $R$ if the energy in the mode is
substantial. The same is true for the combination $f_p$ and
$\tau_p$, as well as any combination involving the $w$-mode data.
Still, we should not discard the possibility that there will be
unique events for which $w$-modes carry the bulk of the
energy, as for example the scattering of \cite{TSM99} 
or the infall of a smaller mass on neutron stars \cite{Borelli}.
In such cases, the strategy proposed by Andersson and
Kokkotas \shortcite{ak98} will be useful.

From the data in Tables~\ref{tab1}--\ref{tab3}, it seems 
natural to use the $p$-mode in any scheme for deducing
the stellar parameters. But, as we have already mentioned, there
does not exist a ``nice'' relation between frequency and stellar 
parameters for the $p$-mode, that is independent of the equation of state. Still,
the strategy that would seem the most promising on the basis of
the present results is based on using the frequencies of $f$-
and $p$-modes. A possible method is based on the following steps: The
first step is to invert the empirical relation for the
$f$-mode frequency, Eq.~(\ref{1}), in order to transform the
measurement of $f_f$ to an  estimate of the mean density of the
star. Let us define a parameter $x=({\bar M}/{\bar R}^3)^{1/2}$,
where ${\bar M} \equiv M/1.4 M_{\odot}$ and ${\bar R} \equiv R/10
\;{\rm km}$, in order to simplify the notation. Then, once $f_f$
has been measured $x$ can be computed from
\begin{equation}
x={ {f_f ({\rm kHz}) - 0.78 }\over 1.63 }\;,
\label{x=}
\end{equation}
with a corresponding relative error
\begin{equation}
{\sigma_{x} \over x } = \left[
\left({0.97 \over \rho} {{\rm 1 ms}\over \tau_f}
{1 \over f_{f}-0.78}\right)^2  \\
+ \left({0.01 \over f_{f}-0.78}\right)^2
 + 0.006^2  \right]^{1/2} \;,
\label{errorx}
\end{equation}
where $f_f$ is expressed in kHz.

The first of the three terms in the
square root comes, obviously, from the measurement error of $f_f$
(the extra complicating factor, related with the $Q$ of the $f$-mode, that
arises there when one tries to express the result with respect to the
signal-to-noise, has been simplified to unity since the $Q$ of the 
$f$-mode is generally a large number).
The other two terms arise from dispersion of the data for
the various equation of state. For a typical $f$-mode frequency of 2~kHz, and
damping time 0.2~s the relative error in $x$ is $\sim 0.01$
(assuming $\rho>6$). Actually, the first term in
Eq.~(\ref{errorx}) is negligible and could be omitted. Therefore
\begin{equation}
{\sigma_{x} \over x } \simeq \sqrt{
\left({0.01 \over f_{f}-0.78}\right)^2
 + 0.006^2                          } \;.
\label{errorxx}
\end{equation}
Here, we need not worry about the sign of the factor $f_f
-0.78$; it is always positive since  $f_f > 1.4\;{\rm kHz}$
for every stellar model in the dataset, see Figure~\ref{fig1} here, or Figure~1
of \cite{ak98}.

\begin{figure}
\centerline{\epsfxsize=8cm \epsfysize=8cm \epsfbox{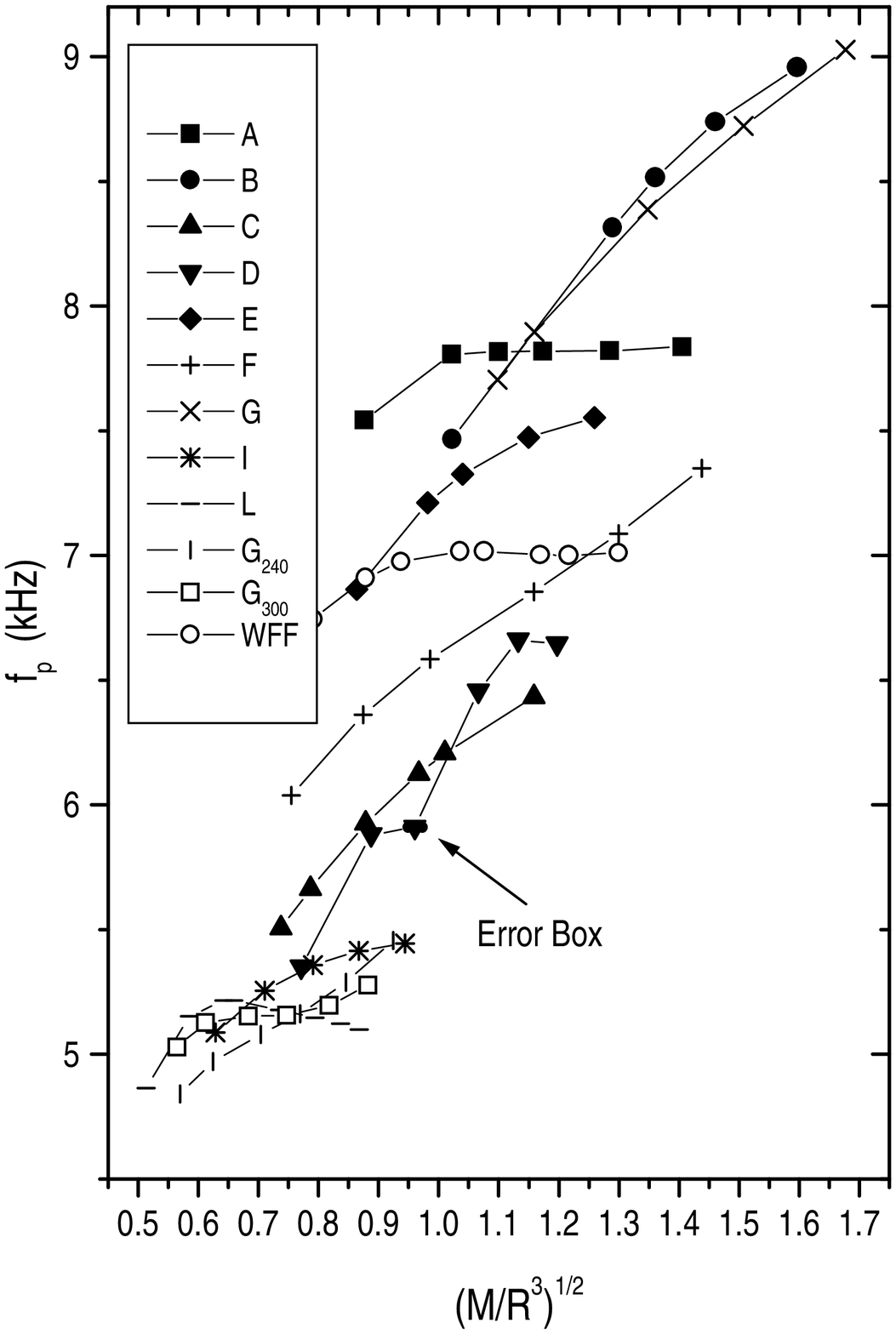}}
\caption{ This diagram shows how the observable frequency of the
first $p$-mode scales with the square root of the mean density of
a neutron star, (${\bar M} / {\bar R}^3$), according to numerical
data for various equations of state. } \label{fig3}
\end{figure}

Then, by measuring the frequency of the first $p$-mode ---which
is expected to carry roughly as much energy as the $f$-mode, see
Allen et al \shortcite{aaks}, one could place an error box on a
diagram of $f_p$ vs. $x$, where all theoretical models for the
equation of state are drawn. We illustrate this in
Figure~\ref{fig3}. This way we can identify the most likely
equation of state. Detecting gravitational waves from pulsating
neutron stars ensures quite accurate measurements of
$f_f$ and $f_p$, so the error box in
Figure~\ref{fig3} is remarkably small. Hence, this method can easily
distinguish between the different equations of state in our
dataset.  In fact, what makes this method so efficient is the fact
that different equations of state are described by quite distinct curves
in an $f_p-x$ diagram.

\begin{figure}
\centerline{\epsfxsize=8cm \epsfysize=8cm \epsfbox{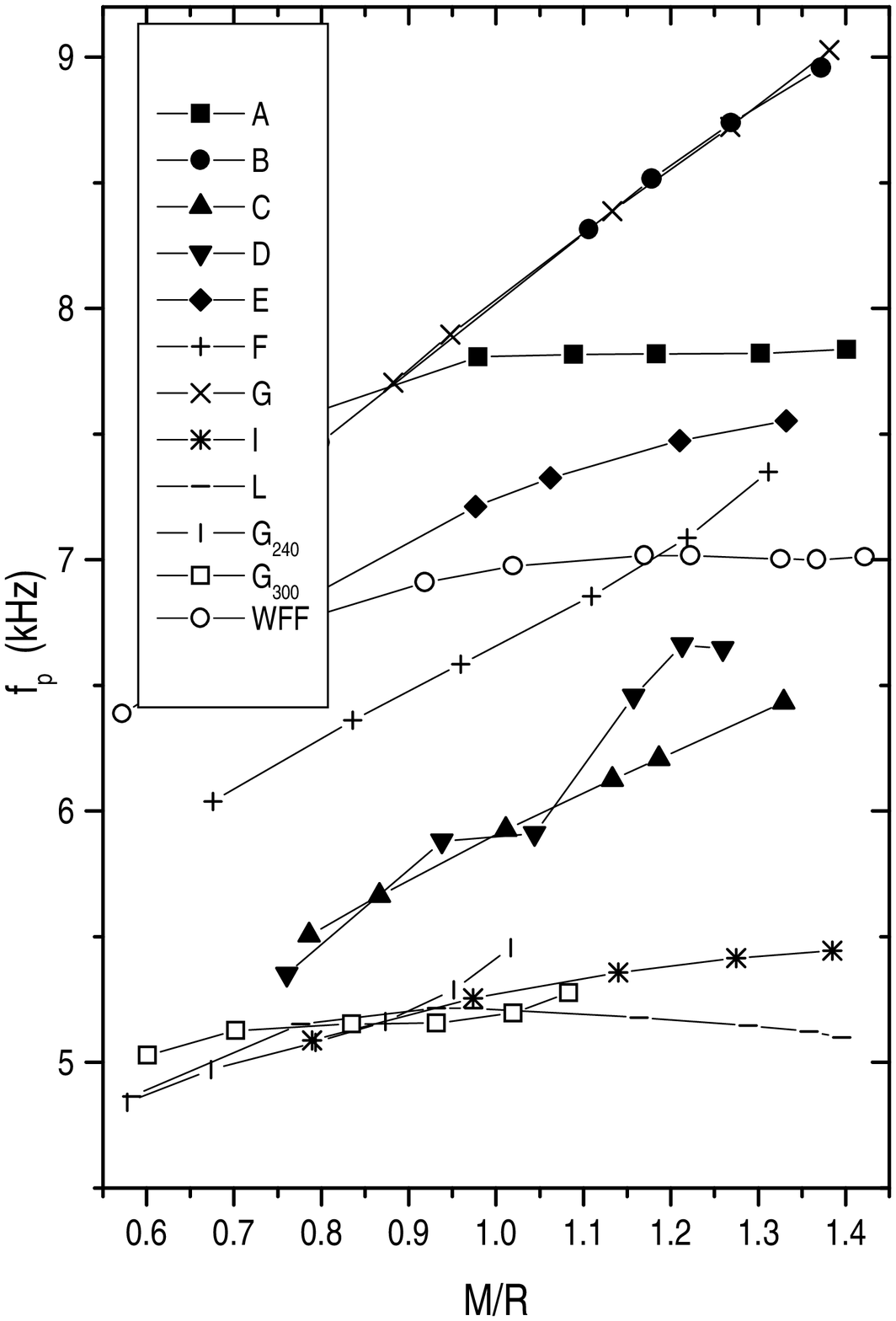}}
\caption{ This diagram shows how the observable frequency of the
first $p$-mode scales with the compactness of a neutron star,
(${\bar M} / {\bar R}$), according to numerical data for
various equations of state. } \label{fig4}
\end{figure}

Finally, we would like to infer the mass and radius of the neutron
star. To do this we can use the data in Figure~\ref{fig4}, which
gives the connection between the $p$-mode frequencies of the
different equations of state and the stellar compactness. From
this diagram, we can immediately use the detected $f_p$ to infer
the compactness of the star, once the right equation of state has
been identified. Having estimated both the
average density and the compactness, it is an elementary
calculation to obtain the mass and the radius of the star.

At this point it is natural to ask the following question: What
if the true equation of state is not close to one of the present
models? This may well be the case. After all, our understanding of
the state of nuclear matter at supranuclear densities still
awaits observational testing.  Should the equation of
state be markedly different from the ones in our sample we will
get an error box which does not lie close to any of the equations
of state in Figure~\ref{fig3}. Should one then want to estimate
the stellar parameters, one could construct a
sequence of polytropes ($p=K\rho^{\gamma}$, $\gamma=1+1/n$) for
various values of $K$ and $n$ (see for example the graphs in
\cite{ak96}). The appropriate combination of these free parameters
will bring the corresponding equation of state within our error
box. Then, one could use Figure~\ref{fig4} and read off the ``correct''
compactness of the star (${\bar M}/{\bar R}$) and from this
compute its radius and mass.

\section{Determining the position of the source}

As with other kinds of gravitational-wave sources, a network of at
least three detectors is needed to pinpoint the location of the
source in the sky. The difference in arrival time for the three
 detectors could be used to determine the position of the
source. The higher the accuracy in measuring the time of arrival
at each detector, the more precise will be the positioning of the
source. Two remote detectors, at a distance $d$ apart from each
other will receive the signal with a temporal difference of
\begin{equation}
\Delta T={d \over c} \cos\theta \;,
\label{deltaT}
\end{equation}
where $c$ is the speed of light, and $\theta$ is the angle between
the line joining the two detectors and the line of sight of the source.
Therefore, the accuracy by which this angle can be measured is
\begin{equation}
\Delta \theta={\sqrt{2} \sigma_{T} c \over d \sin\theta}\;.
\label{deltatheta}
\end{equation}
The $\sqrt{2}$ arises from the measurement errors of the two times
of arrival. If one assumes an `L' shaped network of 3 detectors
with arm length of $d= 10,000$~km, Eqs.~(\ref{relerrortime}) and
(\ref{deltatheta}) lead to an error box on the sky with angular
sides of $1^{\rm o}$, at most (for specific areas of the sky, and large
signal-to-noise ratios the angular sizes could be much smaller).
This is quite interesting since one could then correlate
the detection of gravitational waves with radio- , X-ray or gamma-ray
observations directed on that specific corner of the sky.

\section*{CONCLUDING REMARKS}

In this paper we have extended previous studies of the detectability of
gravitational waves from pulsating neutron stars to include the statistical
errors associated with an analysis of a weak signal in a noisy data stream. 
We have shown that the generation of detectors that is presently under
construction may well be able to observe such sources from within our own
galaxy. Detections from distant galaxies seem unlikely, unless a sizeable
amount of energy (of the order of $0.01M_\odot c^2$) is released
through the pulsation modes. This means that we expect the event
rate to be rather low. One would
certainly not expect to see more than perhaps three neutron stars being born
in supernovae per century in our galaxy.
Of course, other unique 
events in a neutron stars life, like starquakes and phase-transitions, may lead
to relevant signals. Perhaps the most interesting possibility is the
possible association between gravitational waves from pulsation modes and
gamma rays from the soft-gamma repeaters (the magnetars). The number of such
events is very hard to estimate at the present time. Given the many
uncertainties in the available models of all these scenarios,  we certainly
cannot rule out the possibility of detecting   gravitational waves from
pulsations in neutron stars.

However, and this is an important point,
the chances of detecting  pulsating neutron stars would be much enhanced
if one could construct dedicated high-frequency gravitational-wave detectors
sensitive in the range 5-10~kHz. This provides a serious
challenge to experimenters, but that the pay-off of a successful detection of
oscillating  neutron stars would be great
is clear from our present analysis of the inverse problem. 
We have shown that the oscillation frequencies of the fluid $f$ and $p$
modes can be accurately deduced from detected
signals. Other parameters, like
the particulars of the gravitational-wave $w$ modes will be much
more difficult to infer, unless the signal-to-noise ratio of the detection is
unexpectedly large. This means that a previously proposed strategy 
for deducing the parameters of the star (the mass and the radius) from
observations of the $f$ and a $w$-mode is unlikely to be practical.
However, we show that an equally powerful strategy can be based on detected
$f$ and $p$-modes. 
Given an observation of $f$ and $p$-mode oscillation
frequencies, with the estimated accuracies, we can easily 
rule out many proposed equations of state. In other words, we have
proposed a gravitational-wave ``fingerprint analysis'' for neutron stars 
holds a lot of promise, and
may help us reveal the true equation of state at supranuclear densities
once gravitational waves are observed.

\section*{ACKNOWLEDGMENTS}

We are grateful to Bernard Schutz for many illuminating discussions. 
KDK would like to thank British Counsil for a travel grant.



\appendix

\section{EMPIRICAL RELATIONS}

In this Appendix we list the empirical relations deduced from data for
twelve different realistic equations of star. These relations
between the observables (frequency and damping of the modes) and
the stellar parameters (mass $M$ and radius $R$) are essentially
the same as the ones listed by Andersson and Kokkotas
\shortcite{ak98}, but here we also include the relevant
statistical errors.

We have constructed empirical relations for both the
fundamental mode of fluid pulsation (the $f$-mode) and the
slowest damped of the gravitational wave $w$-modes.
For the $f$-mode,  the frequency scales with
the mean density of the neutron star according to
\begin{equation}
{ f_f \over 1 \; {\rm kHz} } \simeq
(0.78  \pm 0.01 ) + (1.63 \pm 0.01)
\left( {{\bar M} \over {\bar R}^3} \right)^{1/2} \;,
\label{1}
\end{equation}
while the damping time of the $f$-mode can be described by
\begin{equation}
{1  {\rm s} \over \tau_f} \simeq
\left(
{{\bar M}^3 \over {\bar R}^4}
\right) \left[
(22.85 \pm 1.51) -
(14.65 \pm 1.32)
\left( {{\bar M} \over {\bar R}} \right)
 \right] \;,
\label{2}
\end{equation}
where ${\bar M} \equiv M/1.4 M_{\odot}$ and ${\bar R}
\equiv R/10 \;{\rm km}$.
The indicated uncertainties are simply the statistical
errors that show up if one takes into account all
stellar models in the data set, on equal basis.
In principle, equations (\ref{1}) and (\ref{2})
could be inverted to compute the
average density (${\bar M}/{\bar R}^3$) and
compactness (${\bar M}/{\bar R}$)
of the star from measured values of $f_f$ and $\tau_f$. However,
this procedure proves unreliable since Eq.~(\ref{2}) is a double-valued
function with respect to compactness, and the presence of errors makes it
impossible to infer the compactness of the star with reasonable accuracy,
even if the characteristics of the waves could be measured with extreme
precision.

The $w$-modes  are pulsations directly associated with spacetime itself.
They have relatively high oscillation frequencies
(6-14 kHz for typical neutron stars)  and barely excite any fluid motion.
They are also rapidly damped, with typical lifetime of fraction
of a  millisecond. The frequency of the
first $w$-mode scales with the stellar
compactness as
\begin{equation}
{ f_w \over 1 \; {\rm kHz} } \simeq
{1 \over {\bar R}}
\left[
(20.95 \pm 0.33) - (9.17 \pm 0.29)
\left( {{\bar M} \over {\bar R}} \right)
\right] \;,
\label{3}
\end{equation}
while the damping rate of the mode is well described by 

\begin{eqnarray}
{ {{\bar M}} \over \tau_w \mbox{ (ms)} } &\simeq &
(3.90 \pm 4.39) + (104.06 \pm 8.33)\left( {{\bar M} \over {\bar R}} \right) \cr
&-& (67.28 \pm 3.84)\left( {{\bar M} \over {\bar R}} \right)^2 
 \ .
\label{4}
\end{eqnarray}

\section{THE FISHER AND COVARIANCE MATRIX}

For a typical mode signal the components
of the symmetric Fisher matrix, defined by
\begin{equation}
\Gamma_{i j} \equiv
2 \Big\langle
{\partial h \over \partial \theta_i} \mid
{\partial h \over \partial \theta_j}
\Big\rangle \;,
\label{Fisher}
\end{equation}
where $\theta_i=(\epsilon, \eta,\zeta,\xi)$
are
\begin{equation}
\Gamma_{\epsilon \epsilon} =
{1 + 24 Q^4 + 32 Q^6 \over (1 + 4 Q^2)^2} \rho^2 \;,
\label{fisher11}
\end{equation}
\begin{equation}
\Gamma_{\epsilon \eta} =
{3 - 4 Q^2 \over 2 (1 + 4 Q^2)^2} \rho^2 \;,
\end{equation}
\begin{equation}
\Gamma_{\epsilon \zeta}=
-2 Q^2 \rho^2 \;,
\end{equation}
\begin{equation}
\Gamma_{\epsilon \xi}=
{1 \over 1 + 4 Q^2} \rho^2 \;,
\end{equation}
\begin{equation}
\Gamma_{\eta \eta}=
{3 + 6 Q^2 + 8 Q^4 \over (1 + 4 Q^2)^2} \rho^2 \;,
\label{fisher22}
\end{equation}
\begin{equation}
\Gamma_{\eta \zeta}=
{1 \over 2} \rho^2 \;,
\end{equation}
\begin{equation}
\Gamma_{\eta \xi}=
{3 + 4 Q^2 \over 2 (1 + 4 Q^2)} \rho^2 \;,
\end{equation}
\begin{equation}
\Gamma_{\zeta \zeta}=
(1 + 4 Q^2) \rho^2 \;,
\end{equation}
\begin{equation}
\Gamma_{\zeta \xi}=
0 \;,
\end{equation}
\begin{equation}
\Gamma_{\xi \xi}=
\rho^2 \;.
\end{equation}

The inverse of the Fisher matrix, $\Sigma_{i j} \equiv \Gamma_{i j}^{-1}$,
the so-called covariance matrix, has the following components:
\begin{equation}
\Sigma_{\epsilon \epsilon}=
{1 - 2 Q^2 + 8 Q^4 \over 2 Q^4 (1 + 4 Q^2)} {1 \over \rho^2} \;,
\label{sigma11}
\end{equation}
\begin{equation}
\Sigma_{\epsilon \eta} =
{3 - 4 Q^2 \over Q^2 (1 + 4 Q^2)} {1 \over \rho^2} \;,
\end{equation}
\begin{equation}
\Sigma_{\epsilon \zeta}=
{-1 + 4 Q^2 \over 2 Q^2 (1 + 4 Q^2)} {1 \over \rho^2} \;,
\end{equation}
\begin{equation}
\Sigma_{\epsilon \xi}=
{-1 + Q^2 \over 2 Q^4} {1 \over \rho^2} \;,
\end{equation}
\begin{equation}
\Sigma_{\eta \eta}=
{4 (5 + 4 Q^2) \over (1 + 4 Q^2)} {1 \over \rho^2} \;,
\label{sigma22}
\end{equation}
\begin{equation}
\Sigma_{\eta \zeta}=
{-4 \over (1 + 4 Q^2)} {1 \over \rho^2} \;,
\end{equation}
\begin{equation}
\Sigma_{\eta \xi}=
{-3 - 2 Q^2 \over Q^2} {1 \over \rho^2} \;,
\end{equation}
\begin{equation}
\Sigma_{\zeta \zeta}=
{2 \over (1 + 4 Q^2)} {1 \over \rho^2} \;,
\label{sigma33}
\end{equation}
\begin{equation}
\Sigma_{\zeta \xi}=
{1 \over 2 Q^2} {1 \over \rho^2} \;,
\end{equation}
\begin{equation}
\Sigma_{\xi \xi}=
{(1 + 2 Q^2)^2 \over 2 Q^4} {1 \over \rho^2} \;.
\label{sigma44}
\end{equation}
These relations can be used to estimate the measurement error in
the various parameters of the mode-signal; see discussion in
the main text.

\end{document}